\documentclass[12pt]{article}
\usepackage[russian]{babel}
\title{The origin of mass and the experiments on future high energy accelerators.}
\author{B.L.Ioffe\\
 Institute of Theoretical
and Experimental Physics,\\ B.Cheremushkinskaya 25, 117218
Moscow,Russia}
\date{}
\begin{document}
\maketitle

\newcommand{\be}{\begin{equation}}
\newcommand{\ee}{\end{equation}}

\def\la{\mathrel{\mathpalette\fun <}}
\def\ga{\mathrel{\mathpalette\fun >}}
\def\fun#1#2{\lower3.6pt\vbox{\baselineskip0pt\lineskip.9pt
\ialign{$\mathsurround=0pt#1\hfil##\hfil$\crcr#2\crcr\sim\crcr}}}

\begin{center}
{\bf Abstract}
\end{center}

\vspace{5mm}

The visible universe  -- it is the universe of nucleons and
electrons. The appearance of nucleon mass is caused by the
violation of chiral symmetry in quantum chromodynamics (QCD). For
this reason, the experiments on high energy accelerators cannot
shed light on the origin of the matter in the visible universe.
The origin of the mass of matter will be clarified, when the
mechanism of chiral symmetry violation in QCD will be elucidated.

\vspace{5mm}

PACS numbers: 12.15. Ff,~ 12.38 Aw, 14.80. Bn

\vspace{5mm}

The Large Hadronic Collider (LHC) with $7 \times 7$ TeV proton
colliding beams is now under construction at CERN. It is planned,
that the accelerator will be put into operation in 2007, the
experiments will start in 2008. The possibilities of construction
of other high energy accelerators are under discussion: the linear
accelerator with colliding $e^+e^-$ beams of energies 0.5 - 1.0
TeV, muon and proton colliders. The goal of experiments on future
high energy accelerators is to investigate physics beyond the
Standard Model, i.e, the study of processes at energies
inaccessible on existing accelerators, the search for particles,
which were not found till now, but expected: Higgs bosons,
supersymmetric particles etc. The Standard Model is the
$SU(3)\times SU(2)\times (U(1)$ gauge theory, where $SU(3)$
corresponds to quantum chromodynmics -- the theory of strong
interactions, $SU(2)\times (U(1)$ -- to electroweak theory, which
unites the electromagnetic and weak interactions. The experiments,
performed on high energy accelerators, which are (or were) in
operation, agree with the Standard Model and give no indications
for the presence of any new physics beyond it. The discovery of
neutrino oscillations, the first data on the existence of massive
majorano neutrino, as well as the astrophysical data -- the
existence of dark matter -- indicate the evidence for such new
physics. For this reason, great expectations are due to
experiments on LHC and other future accelerators.

One of the most important problems, unsolved in the course of the
investigation of the Standard Model, the problem, which, as it is
expected, will be solved by LHC experiments (and other future
accelerators) is the problem of the origin of particle masses. As
it was mentioned in the report by De Roeck, J.Ellis and F.Gianotti
\cite{1}, devoted to experiments on future accelerators, "The most
prominent areas, where the Standard Model leaves unsolved
problems, include the origin of particle masses ..." . In the
joint proposal of experiments on LHC, presented by ATLAS and CMS
collaborations \cite{2} it said: "The $SU(3)\times SU(2)\times
U(1)$ gauge interactions of the Standard Model ... give no
explanation of the origin of particle masses". The main problem of
the investigations on the linear $e^+e^-$ collider is: "The
mecanism of elctroweak symmetry breaking, in other words, what is
the origin of mass in particle physics" \cite{3}. This problem is
formulated in a similar way in other papers and reports devoted to
experiments on future accelerators (see, e.g. \cite{4}, \cite{5}).

The statements, presented above, are based on the initial idea of
the Standard Model. The violation of the gauge SU(2) symmetry,
which is needed in the Standard Model in order to get masses to
$W$ and $Z$ intermediate bosons, is realized by introducing the
scalar field $\varphi$, the SU(2) doublet. The scalar field
$\varphi$ interacts with the fields of $W$, $Z$ quarks, leptons
and itself. It is supposed, that its self-action potential has a
minimum at non-zero value $\varphi = \varphi_0$, which is constant
in space and time. The field $\varphi$ is represented as the
expansion near the minimum of the potential $\varphi = \varphi_0 +
\chi$, where $\chi$ is the quantized field, the quants of which
are Higgs bosons. The masses of intermediate bosons, quarks and
leptons arise due to the constant scalar field $\varphi_0$ because
of the presence of the terms $g^2_1 \varphi^2_0 W^2$, $g^2_2
\varphi^2_0 Z^2$, ~~ $h_{l i} \varphi_0 \bar{l}_i l_i$,~~ $h_{q i}
\varphi_0 \bar{q}_i q_i$ in the interaction Lagrangian. (Here
$l_i, q_i$ are the fields of leptons and quarks, $h_{li},h_{qi}$
are the coupling constants). Therefore, the observation of Higgs
boson and its interactions with quarks and leptons will explain
the origin of quark and lepton masses.

Just such a possibility was taken in mind in the cited above
statements.

The goal of this paper is to pay attention on the fact, that
elucidation of the problem of Higgs boson existence and
clarification of the structure of its interactions with quarks
(and even more generally, the study of interactions at high
energies) has no relation (more precisely, has very little
relation) to the origin of mass of particles of visible universe.
The visible world - stars, planets, galactics, all bodies
surrounding us, are built from protons, nuclei and electrons. The
electrons contributions to the mass of the matter is negligibly
small -- less  than 0.1 \%. Therefore, in order to understand the
origin of the mass of the observable world, it is necessary to
clarify the origin of nucleon mass. (I leave temporarily aside the
problem of the mass of dark matter). The nucleon consists of $u$-
and $d$-quarks. But the masses of $u$- and $d$-quarks are small,
$m_u + m_d \approx 10~MeV$ and their mass contributes a small
part, about 1-2 \% to the nucleon mass. It can be shown, that in
the formal limit $m_u, m_d \to 0$, the nucleon mass practically
does not change. The nucleon mass arises  due to spontaneous
violation of chiral symmetry in QCD  and can be expressed through
the chiral symmetry violating vacuum condensates \cite{6}. There
is an approximate formula, which expresses the nucleon mass
through quark condensate \cite{6}, \cite{7}, \cite{8}:
\be m = \Biggl [ - 2(2 \pi)^2 \langle 0 \vert \bar{q} q \vert 0
\rangle \Biggr ] ^{1/3} \label{1}
\ee

 Here $m$ - is the nucleon mass, $\langle 0 \vert \bar{q}q \vert
0 \rangle$ is the quark condensate, $q$ is the field of $u$- or
$d$-quarks. (The recent reviews of the modern status of chiral
symmetry in QCD, its violation and the problem of proton mass were
given in \cite{9}, \cite{10}.) The expression for quark condensate
(the Gell-Mann-Oakes-Renner formula)

\be \langle 0 \vert \bar{q} q \vert 0 \rangle = -\frac{1}{2} ~
\frac{m^2_{\pi} f^2_{\pi}}{m_u + m_d} \label{2} \ee follows from
chiral symmetry in QCD (conservation of the axial current violated
by quark masses). Here $m_{\pi}$ and $f_{\pi}$ are the pion mass
and decay constant. In the limit of  quark masses going to zero,
$m_u, m_d \to 0$~ $m^2_{\pi}$ is also going to zero in such a way
that $\langle 0 \vert \bar{q} q \vert 0 \rangle$ is a constant in
this limit. The recent determination of the value of quark
condensate gave \cite{10} $\langle 0 \vert \bar{q} q \vert o
\rangle = - (254 MeV)^3 \pm 10 \%$. The substitution of this value
into (\ref{1}) results in $m = 1.09 GeV$ in comparison with the
experimental value $m = 0.94 GeV$. Therefore, in order to clear up
the origin of nucleon mass, it is necessary to clarify the
mechanism of chiral symmetry violation in QCD. The chiral symmetry
violation proceeds at large distances $\sim 1 fermi = 10^{-13}cm$
and is not related to experiments at high energies, where the
domain of small distances $\sim 10^{-17}cm$ is investigated.
Clarification of the mechanism of chiral symmetry violation in QCD
-- is the theoretical problem. Perhaps, lattice calculation would
be helpful here, as well as the experiments on heavy ions
collisions at high energies, from which  some information on phase
transitions in QCDD could be extracted. However, the baryon
density is high in such experiments, i.e., the chemical potential
$\mu$ is large, whereas in vacuum $\mu = 0$. The phase transition
at finite $\mu$ and $\mu = 0$ could be quite different \cite{12},
\cite{13}.

The mass of the electron arises in the Standard Model due to
interaction with the Higgs field. Therefore, the origin of the
electron and the proton mass is quite different and their strong
distinction is not surprising.

A short remark about dark matter. Its nature is unknown. Perhaps,
dark matter particles are very heavy and weakly interact with
usual matter. In this case it is unprobably, that their mass
arises because of interaction with the Higgs field.

I am thankful too A.V.Samsonov for collection of the material on
planned experiments on future accelerators. This work was
supported in part by CRDF Cooperative Grant Program, Project
RUP2-2621-MO-04 and the funds from EC to the project "Study of
Strongly Interacting Matter" under contract 2004 No
R113-CT-22004-506078.

\end{document}